# Analog Beamsteering for Flexible Hybrid Beamforming Design in Mmwave Communications

Yaning Zou, Wolfgang Rave and Gerhard Fettweis
Vodafone Chair Technische Universität Dresden
email: {yaning.zou, rave, gerhard.fettweis}@ifn.et.tu-dresden.de

*Abstract* — In this paper, we propose an analog beamsteering approach for enabling flexible hybrid beamforming design that can achieve performance close to singular value decomposition (SVD) based digital beamforming with single user case. As a starting point, assuming the use of a codebook with infinite precision, we apply transposes of antenna array response matrices as analog beamformers at the BS and the UE sides. The resulting effective channel matrix including both propagation channel and analog beamformers is shown to be able to provide a maximal achievable rate very close to SVD based digital beamforming at low to medium SNR. Next, the impact of a finite size codebook is analyzed for practical analog beamformer design. A closed-form derivation is obtained for mapping rate loss to the number of antennas, codebook sizes and the received SNR. The analysis shows that in order to achieve performance comparable to analog beamforming with infinite precision, the codebook size should be at least larger than the number of implemented antennas. Therefore, the proposed analog beamforming approach can be designed independently without taking digital beamforming into account. Such an approach can lead to the development of flexible hybrid beamforming architectures that could adapt to a changing propagation environment via agile analog beam selection and maximize spatial multiplexing gain via proper digital beamformer design.

## I. INTRODUCTION

One important aspect for the development of 5G and future communication evolution is to provide extreme mobile broadband user experiences, such as UHD/3D streaming, immersive applications and ultra-responsive cloud services, with an affordable price [1]-[3]. In addition to improve spectrum usage at the 1-6 GHz frequencies, mobile communication systems operating in mmwave bands are being seriously considered by the industry as a very promising approach to boost capacity significantly [1]-[3]. By deploying a large number of antennas at the BS and UE sides, severe path loss problem that traditionally limits scope of communication over mmwave frequency bands can be addressed [1], [3], [4].

However, in practice, it is very challenging to devise beamforming techniques that are capable of quickly adapting to constantly changing propagation environment in mmwave frequency bands and at the same time provide extremely high throughput with reasonable cost and energy efficiency. Based on measurement results shown in [4], in addition to the traditionally considered line-of-sight (LOS) path, one or multiple strong reflecting paths can also exist in the mmwave propagation environment. In order to capture spatial multiplexing gain offered by the multi-path phenomenon with reasonable implementation cost, several hybrid beamforming techniques are proposed in [5]-[7]. The main idea is to optimize analog and digital beamformers together at the BS and UE side, separately or jointly, for achieving the best possible performance. Yet such a digital/analog co-optimization approach might not be the best candidate for adapting fast varying mmwave propagation environment as it takes some computation resources and time to carry out optimization process even if perfect channel knowledge is already acquired.

In this paper, we propose an array response vector based analog beamsteering approach that can lead to flexible hybrid beamforming design with performance close to SVD based digital beamforming. In detail, assuming the use of a codebook with infinite precision, we apply transposes of antenna array response matrices as analog beamformers at the BS and the UE sides. The resulting effective channel matrix including both propagation channel and analog beamformers can provide maximal achievable rate very close to SVD based digital beamforming at lower to medium SNR. Next, the impact of finite codebook size is analyzed in a closed-form. The obtained derivation can evaluate rate loss mathematically, in terms of the number of antennas, codebook sizes and the received SNR, without using extensive computer simulations. Based on the analysis, the codebook size should be at least larger than the number of implemented antenna elements. In order to minimize performance degradation, the codebook size should be about 2 times of the number of implemented antennas.

*Notations:* $\text{diag}(\mathbf{a})$ denotes a square diagonal matrix with elements of the vector $\mathbf{a}$ on the diagonal. Superscripts $(.)^T$ and $(.)^H$ denote vector or matrix transpose and vector or matrix conjugate transpose, respectively. Finally, $(.)^{-1}$ denotes matrix inverse.

## II. ANALOG BEAMSTEERING BASED ON ARRAY REPONSE VECTORS USING A CODEBOOK WITH INFINITE PRECISION

### A. Basic Signal Model for Hybrid Beamforming

We consider a single-user point-to-point DL transmission system that operates at mmwave frequencies. At the BS side, $N_{BS}$ transmit antennas and $M_{BS}$ transceivers are implemented, while at the UE side, $N_{UE}$ receive antennas and $M_{UE}$ transceivers are deployed.

As discussed in [4], [6], the propagation channel at the mmwave frequencies has a limited number of scatterers. Assume the transmitted signal is narrowband compared to the operating central frequency. We can then write the $N_{UE} \times N_{BS}$ propagation channel matrix using a geometric model with $L$ scatterers [6], [8] as

$$\mathbf{H} = \mathbf{A}_{UE} \text{diag}(\mathbf{g}) \mathbf{A}_{BS}^H \quad (1)$$

where $\mathbf{g} = \sqrt{N_{UE} N_{BS} / \rho}\, [g_1, g_2, ..., g_L]$, $\rho$ refers to the path-loss between the BS and the UE, $g_l$ denotes a complex gain of the $l$-th path and follows a complex Gaussian distribution as $g_l \sim \mathbb{CN}(0, \sigma_l^2)$. Assuming uniform linear array (ULA) is implemented at the both sides, the antenna array response matrices at the BS and the UE read

$$\begin{aligned}\mathbf{A}_{BS} &= [\mathbf{a}_{BS}(\phi_1), ..., \mathbf{a}_{BS}(\phi_L)] \\ \mathbf{A}_{UE} &= [\mathbf{a}_{UE}(\theta_1), ..., \mathbf{a}_{UE}(\theta_L)]\end{aligned} \quad (2)$$

respectively, where

$$\mathbf{a}_{BS}(\phi_l) = \frac{1}{\sqrt{N_{BS}}}\left[1, e^{j(2\pi/\lambda)d\sin(\phi_l)}, ..., e^{j(N_{BS}-1)(2\pi/\lambda)d\sin(\phi_l)}\right]^T$$

$$\mathbf{a}_{UE}(\theta_l) = \frac{1}{\sqrt{N_{UE}}}\left[1, e^{j(2\pi/\lambda)d\sin(\theta_l)}, ..., e^{j(N_{UE}-1)(2\pi/\lambda)d\sin(\theta_l)}\right]^T$$

Here, $\phi_l$ and $\theta_l$ refer to azimuth angles of departure (AoD) at the BS and azimuth angles of arrival (AoA) at the UE, respectively. Both $\phi_l$ and $\theta_l$ follow uniform distribution over $[-0.5\pi, 0.5\pi]$. $\lambda$ is the wavelength of transmitted signal and $d$ denotes distance between antenna elements. The antenna array response vector can also be written in terms of wavenumber as $\mathbf{a}_{BS}(\kappa_{l,BS})$ and $\mathbf{a}_{UE}(\kappa_{l,UE})$ where $\kappa_{l,BS} = 2\pi \sin\phi_l / \lambda$ and $\kappa_{l,UE} = 2\pi \sin\theta_l / \lambda$.

In theory, this channel matrix described in (1) can support up to $L$ independent data streams. Here, we exploit spatial multiplexing gain offered by the channel and transmit $L \times 1$ data vector $\mathbf{s}$ from the BS. After experience digital/analog beamforming at the both BS and UE sides, signal vector at the detector input at the UE side appears to be

$$\mathbf{y} = \mathbf{W}_D \mathbf{W}_A \mathbf{H} \mathbf{F}_A \mathbf{F}_D \mathbf{s} + \mathbf{W}_D \mathbf{W}_A \mathbf{n} \quad (3)$$

where $\mathbf{F}_A$ and $\mathbf{F}_D$ are $N_{BS} \times M_{BS}$ analog and $M_{BS} \times L$ digital beamformers at the BS side and $\mathbf{W}_A$ and $\mathbf{W}_D$ refer to $M_{UE} \times N_{UE}$ analog and $L \times M_{UE}$ digital beamformers at the UE side, respectively, and $\mathbf{n}$ denotes $N_{UE} \times 1$ additive channel noise vector with complex Gaussian distributed entries following $\mathbb{CN}(0, \sigma_n^2 / N_{UE})$. The SNR is defined as $\chi = P_s \sum_{l=1}^L \sigma_l^2 / (\rho \sigma_n^2)$ where $P_s$ denotes the transmitted power of $\mathbf{s}$. The achievable array gain is given by $10\log_{10}(N_{UE} N_{BS})$.

In general, the performance of hybrid beamforming is upper bounded by digital beamforming, e.g., based on SVD decomposition [4]-[7]. As a reference, we consider the widely known optimal unconstrained unitary digital precoder given by $N_{BS} \times L$ unitary matrix $\mathbf{F}_{opt} = \mathbf{F}_A \mathbf{F}_D = \mathbf{V}_L$. It consists of the $L$ right singular vectors of $\mathbf{H}$ based on singular value decomposition (SVD) over the transmission channel matrix $\mathbf{H}$ as $\mathbf{H} = \mathbf{U}\mathbf{\Lambda}\mathbf{V}^H$ [6]. At the UE side, applying $L \times N_{UE}$ combiner $\mathbf{U}_L^H = \mathbf{W}_D \mathbf{W}_A$ that consists of the $L$ left singular vectors of $\mathbf{H}$ at the UE side, the reference maximal achievable rate is given by

$$R_D = E\left\{\log_2\left[\det\left(\mathbf{I} + P_s \frac{\mathbf{\Lambda}_L \mathbf{\Lambda}_L^H}{\sigma_n^2}\right)\right]\right\} \quad (4)$$

where $\mathbf{\Lambda}_L$ is a $L \times L$ diagonal matrix contains entries of singular values of $\mathbf{H}$. The expectation is carried out over different channel realizations.

### B. Analog Beamsteering based on Array Reponse Vectors using A Codebook with Infinite Precision

Assume perfect knowledge on AoD and AoA known at the BS and the UE respectively. Also, assume the analog beamformers use codebooks with infinite precision at the both sides. $L$ transceivers are implemented at the both BS and UE sides also, i.e., $M_{BS} = M_{UE} = L$.

Then we directly apply transposes of the antenna array response matrices $\mathbf{A}_{BS}^H$ and $\mathbf{A}_{UE}$ in (1) as analog beamformers as

$$\mathbf{F}_A = \mathbf{A}_{BS} \quad \mathbf{W}_A = \mathbf{A}_{UE}^H \quad (5)$$

Incorporating (5) with (1), the resulting effective channel matrix including propagation channel matrix and analog beamformers reads

$$\bar{\mathbf{H}} = \mathbf{W}_A \mathbf{H} \mathbf{F}_A = \tilde{\mathbf{W}}_A \text{diag}(\mathbf{g}) \tilde{\mathbf{F}}_A \quad (6)$$

where both $\tilde{\mathbf{F}}_A$ and $\tilde{\mathbf{W}}_A$ are $L \times L$ square matrices, $\tilde{\mathbf{F}}_A = \mathbf{A}_{BS}^H \mathbf{A}_{BS}$ and $\tilde{\mathbf{W}}_A = \mathbf{A}_{UE}^H \mathbf{A}_{UE}$.

With unit transmit power at the BS, the achievable maximal throughput of the effective channel matrix in (6) can be evaluated as

$$\begin{aligned}R_A &= E\left\{\log_2\left[\det\left(\mathbf{I} + P_s \sigma_n^{-2}\left(\mathbf{W}_A^H \mathbf{W}_A\right)^{-1} \bar{\mathbf{H}}\bar{\mathbf{H}}^H\right)\right]\right\} \\ &= E\left\{\log_2\left[\det\left(\mathbf{I} + P_s \sigma_n^{-2}\left(\mathbf{W}_A^H \mathbf{W}_A\right)^{-1} \tilde{\mathbf{W}}_A \text{diag}(\mathbf{g})\right.\right.\right. \\ &\quad \left.\left.\left.\times \tilde{\mathbf{F}}_A\left(\tilde{\mathbf{W}}_A \text{diag}(\mathbf{g}) \tilde{\mathbf{F}}_A\right)^H\right)\right]\right\}\end{aligned} \quad (7)$$

Note that, this rate defined in (7) only affects the achievable performance of the effective channel matrix $\bar{\mathbf{H}}$. Further processing in digital domain via proper design of digital precoder $\mathbf{F}_D$ and combiner $\mathbf{W}_D$ is still required. This is out of the scope of the paper and forms a continuing topic for the future study.

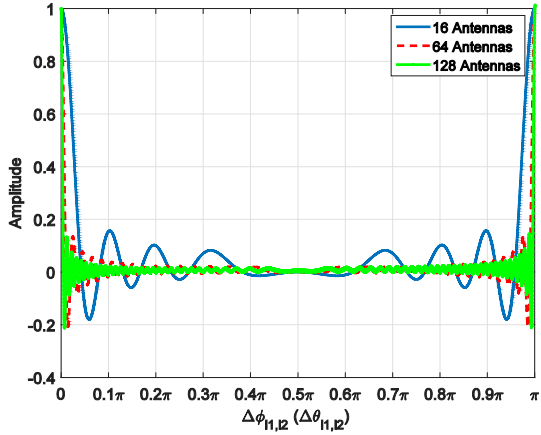

Fig. 1. Example spectral illustrations on amplitude over different phase distance from 0 to $\pi$, $d = \lambda / 2$.

Next, based on (2) and (6), it is interesting to observe that $\tilde{\mathbf{F}}_A$ and $\tilde{\mathbf{W}}_A$ are non-diagonal matrices with all the diagonal entries equal to 1, i.e., $[\tilde{\mathbf{W}}_A]_{l,l} = [\tilde{\mathbf{F}}_A]_{l,l} = 1$ for $l = 1,\ldots,L$. Amplitudes of the corresponding off-diagonal elements at $l_1$-th row and $l_2$-th column read

$$\left| [\tilde{\mathbf{F}}_A]_{\substack{l_1,l_2 \\ l_1 \neq l_2}} \right| = \left| \frac{\sin[\pi d N_{BS}(\sin\phi_{l_2} - \sin\phi_{l_1})/\lambda]}{N_{BS}\sin[\pi d(\sin\phi_{l_2} - \sin\phi_{l_1})/\lambda]} \right|$$
$$\left| [\tilde{\mathbf{W}}_A]_{\substack{l_1,l_2 \\ l_1 \neq l_2}} \right| = \left| \frac{\sin[\pi d N_{UE}(\sin\theta_{l_2} - \sin\theta_{l_1})/\lambda]}{N_{UE}\sin[\pi d(\sin\theta_{l_2} - \sin\theta_{l_1})/\lambda]} \right| \quad (8)$$

In general, the amplitudes of $\left|[\tilde{\mathbf{F}}_A]_{l_1,l_2}\right|$ and $\left|[\tilde{\mathbf{W}}_A]_{l_1,l_2}\right|$ drops dramatically with increased phase difference $\Delta\phi_{l_1,l_2} = |\phi_{l_2} - \phi_{l_1}|$ and $\Delta\theta_{l_1,l_2} = |\theta_{l_2} - \theta_{l_1}|$. For example, as shown in Fig 1, the probabilities of $\left|[\tilde{\mathbf{F}}_A]_{l_1,l_2}\right|^2 > 0.5$ and $\left|[\tilde{\mathbf{W}}_A]_{l_1,l_2}\right|^2 > 0.5$ are less than $1/N_{BS}$ and $1/N_{UE}$ respectively when $\phi_{l_1} = \theta_{l_1} = 0$. If a large number of antennas are deployed, with a high probability, the off-diagonal elements in (8) are quite small values compared to 1. Then we can make approximation as $\tilde{\mathbf{F}}_A \approx \mathbf{I}$ and $\tilde{\mathbf{W}}_A \approx \mathbf{I}$. Inherently, the achievable system throughput in (7) can be simplified to

$$R_A \approx E\left\{\log_2\left[\det\left(\mathbf{I} + P_s \frac{\text{diag}(\mathbf{g})\text{diag}(\mathbf{g}^*)}{\sigma_n^2}\right)\right]\right\} \quad (9)$$

As will be shown in simulations, the above approximation in (9) is a good estimate in low-mid SNR range where additive channel noise is the dominant adverse factor. It tends to overestimate achievable rate when SNR is high and inter-stream interference due to non-zero diagonal elements in $\tilde{\mathbf{F}}_A$ and $\tilde{\mathbf{W}}_A$ becomes to be the dominant adverse factor.

### III. ANALOG BEAMSTEERING BASED ON ARRAY REPONSE VECTORS USING FINITE SIZE CODEBOOKS

In the previous section, we simply assume that the analog beamformers use codebooks with infinite precision. In this section, we consider the realistic case that finite size codebooks are used and then analyze the impact of codebook size on the performance in terms of maximal achievable rate.

Assume phase shifters are used as the implementation of antenna array beamsteering. The used codebooks are $\mathbf{E}_{BS}$ at the BS and $\mathbf{E}_{UE}$ at the UE with sizes of $C_{BS}$ and $C_{UE}$, respectively. Entries of the codebooks correspond to different steering angles $\phi_l$ and $\theta_l$ and inherently different steering phases given by $(2\pi/\lambda)d\sin(\phi_l) = d\kappa_{l,BS}$ and $(2\pi/\lambda)d\sin(\theta_l) = d\kappa_{l,UE}$ at the BS and UE sides, respectively. For implementation simplicity, we assume entries in the codebooks are equally distributed over phase domain $[-(2\pi/\lambda)d, (2\pi/\lambda)d]$ instead of steering angles $\phi_l$ and $\theta_l$. With fixed antenna distance $d$, available wavenumber entries $\bar{\kappa}_{l,BS(UE),n}$ in the codebook are

$$\bar{\kappa}_{l,BS(UE),n} = 4n\pi/\lambda/C_{BS,(UE)} - (2\pi/\lambda) \quad (10)$$

for $n = 0,\ldots,C_{BS(UE)} - 1$. The corresponding entries in terms of steering angles are then given by $\arcsin(\lambda\bar{\kappa}_{l,BS(UE),n}/2/\pi)$ that is not equally distributed over angular domain for $n = 0,\ldots,C_{BS(UE)} - 1$.

Now, with codebooks $\mathbf{E}_{BS}$ and $\mathbf{E}_{UE}$ and known $\kappa_{l,BS(UE)}$ from propagation channel, we construct analog beamformers based on (5) as

$$\hat{\mathbf{F}}_A = [\mathbf{a}_{BS}(\hat{\kappa}_{1,BS}),\ldots,\mathbf{a}_{BS}(\hat{\kappa}_{L,BS})]$$
$$\hat{\mathbf{W}}_A = [\mathbf{a}_{UE}(\hat{\kappa}_{1,UE}),\ldots,\mathbf{a}_{UE}(\hat{\kappa}_{L,UE})]^H \quad (11)$$

where $\hat{\kappa}_{l,BS}$ and $\hat{\kappa}_{l,UE}$ are wavenumbers from codebooks $\mathbf{E}_{BS}$ and $\mathbf{E}_{UE}$ that minimize Euclidian distances of

$$\hat{\kappa}_{l,BS(UE)} = \arg\min_{\bar{\kappa}_{l,BS(UE),n}} \left|\kappa_{l,BS(UE)} - \bar{\kappa}_{l,BS(UE),n}\right|$$
$$\text{s.t.} \quad \bar{\kappa}_{l,BS(UE),n} \in \mathbf{E}_{BS,(UE)} \quad (12)$$

In general, differences between the chosen wavenumbers $\hat{\kappa}_{l,BS(UE)}$ in the used codebook and actual wavenumbers $\kappa_{l,BS(UE)}$ from propagation channel in (1) appear to be

$$\kappa'_{l,BS} = \kappa_{l,BS} - \hat{\kappa}_{l,BS} \approx (2\pi/\lambda)\phi'_l\cos\phi_l$$
$$\kappa'_{l,UE} = \kappa_{l,UE} - \hat{\kappa}_{l,UE} \approx (2\pi/\lambda)\theta'_l\cos\theta_l \quad (13)$$

where $\phi' = \phi - \hat{\phi}$ and $\theta' = \theta - \hat{\theta}$. With given $\phi$ and $\theta$, $\kappa'_{l,BS}$ and $\kappa'_{l,UE}$ follow uniform distribution $\kappa'_{BS} \sim (-2\pi/C_{BS}/\lambda, 2\pi/C_{BS}/\lambda)$ and $\kappa'_{UE} \sim (-2\pi/C_{UE}/\lambda, 2\pi/C_{UE}/\lambda)$. It can be shown that the entries of codebooks are orthogonal so that $\hat{\mathbf{F}}_A^H\hat{\mathbf{F}}_A = \mathbf{I}$ and $\hat{\mathbf{W}}_A\hat{\mathbf{W}}_A^H = \mathbf{I}$.

Now incorporating (11) with (1), after applying analog beamsteering using finite size codebooks, the effective channel matrix becomes to be

$$\hat{\tilde{\mathbf{H}}} = \hat{\mathbf{W}}_A\mathbf{H}\hat{\mathbf{F}}_A = \vec{\mathbf{W}}_A\text{diag}(\mathbf{g})\vec{\mathbf{F}}_A \quad (14)$$

where both $\vec{\mathbf{F}}_A$ and $\vec{\mathbf{W}}_A$ are $L \times L$ square matrices, $\vec{\mathbf{F}}_A = \mathbf{A}_{BS}^H \hat{\mathbf{F}}_A$ and $\vec{\mathbf{W}}_A = \hat{\mathbf{W}}_A \mathbf{A}_{UE}$.

The maximal achievable rate of the effective channel matrix including both propagation channel matrix and analog beamformers based on codebooks $\mathbf{E}_{BS}$ and $\mathbf{E}_{UE}$ reads

$$\hat{R}_A = E\left\{\log_2\left[\det\left(\mathbf{I} + P_s\sigma_n^2\left(\hat{\mathbf{W}}_A\hat{\mathbf{W}}_A^H\right)^{-1}\hat{\bar{\mathbf{H}}}\hat{\bar{\mathbf{H}}}^H\right)\right]\right\}$$
$$= E\left\{\log_2\left[\det\left(\mathbf{I} + \frac{P_s\vec{\mathbf{W}}_A\mathrm{diag}(\mathbf{g})\vec{\mathbf{F}}_A\left(\vec{\mathbf{W}}_A\mathrm{diag}(\mathbf{g})\vec{\mathbf{F}}_A\right)^H}{\sigma_n^2}\right)\right]\right\}$$
(15)

In order to analyze the impact of codebook size on the link performance, similar for deriving (9), we again approximate the off-diagonal elements of $\vec{\mathbf{W}}_A$ and $\vec{\mathbf{F}}_A$ to be small valued compared to the diagonal elements so that $\vec{\mathbf{W}}_A$ and $\vec{\mathbf{F}}_A$ are both diagonal matrices and given by

$$\vec{\mathbf{F}}_A \approx \Delta_{BS} \quad \vec{\mathbf{W}}_A \approx \Delta_{UE} \quad (16)$$

where $\Delta_{BS} = [\Delta_{1,BS},...,\Delta_{L,BS}]$ and $\Delta_{UE} = [\Delta_{1,UE},...,\Delta_{L,UE}]$. $\Delta_{l,BS}$ and $\Delta_{l,UE}$ are $l$-th diagonal elements of $\vec{\mathbf{F}}_A$ and $\vec{\mathbf{W}}_A$ and their amplitudes read

$$\left|\Delta_{l,BS}\right| = \left|\left[\vec{\mathbf{F}}_A\right]_{l,l}\right| = \left|\frac{\sin[dN_{BS}\kappa'_{l,BS}/2]}{N_{BS}\sin[d\kappa'_{l,BS}/2]}\right|$$
$$\left|\Delta_{l,UE}\right| = \left|\left[\vec{\mathbf{W}}_A\right]_{l,l}\right| = \left|\frac{\sin[dN_{UE}\kappa'_{l,UE}/2]}{N_{UE}\sin[d\kappa'_{l,UE}/2]}\right|$$
(17)

where $\kappa'_{l,BS} = 2\pi(\sin\phi_l - \sin\hat{\phi}_l)/\lambda$ and $\kappa'_{l,UE} = 2\pi \times (\sin\theta_l - \sin\hat{\theta}_l)/\lambda$.

Incorporating (16) with (15), the maximal achievable rate can be approximated as

$$\hat{R}_A \approx E\left\{\log_2[\det(\mathbf{I} + \Delta_{UE}\mathrm{diag}(\mathbf{g})\Delta_{BS}\right.$$
$$\left.\times P_s\left(\Delta_{UE}\mathrm{diag}(\mathbf{g})\Delta_{BS}\right)^H/\sigma_n^2)]\right\}$$
$$= E\left[\log_2\left[\prod_{l=1}^L\left(\left|\Delta_{l,BS}^2\Delta_{l,UE}^2 g_l^2\beta\right|/\sigma_n^2 + 1\right)\right]\right] \quad (18)$$
$$= \sum_{l=1}^L E\left[\log_2\left(\left|\Delta_{l,BS}^2\Delta_{l,UE}^2 g_l^2\beta\right|/\sigma_n^2 + 1\right)\right]$$

where $\beta = P_sN_{UE}N_{BS}/\rho$. Subtracting (18) from (7) and applying Taylor expansion $\log(1-x) = -\sum_{n=1}^\infty x^n/n$ for $|x| < 1$, the resulting rate loss due to the use of finite size codebook is obtained as

$$\Delta R_A \approx R_A - \hat{R}_A$$
$$= \frac{1}{\log(2)}\sum_{l=1}^L\sum_{n=1}^\infty E\left[\left(\frac{\left|\beta g_l^2\right|/\sigma_n^2}{\left|\beta g_l^2\right|/\sigma_n^2 + 1}\right)^n\right]E\left[\frac{\left(1-\left|\Delta_{l,BS}^2\Delta_{l,UE}^2\right|\right)^n}{n}\right]$$
$$\approx \frac{1}{\log(2)}\sum_{l=1}^L \underbrace{E\left[\frac{\left|\beta g_l^2\right|/\sigma_n^2}{\left|\beta g_l^2\right|/\sigma_n^2 + 1}\right]}_{R_{0,l}}\underbrace{E\left[\left(1-\left|\Delta_{l,BS}^2\Delta_{l,UE}^2\right|\right)\right]}_{R_{1,l}}$$
(19)

In (19), the approximation is made based on the assumption that wavenumber errors $\kappa'_{l,BS(UE)}$, $l = 1,...,L$, are small.

Here, the rate loss in (19) for the $l$-th path is determined by product of two multiplicative terms $R_{0,l}$ and $R_{1,l}$. The first multiplication term $R_{0,l}$ depends on the $l$-th path gain $g_l$ that follows complex Gaussian distribution $\mathbb{CN}(0,\sigma_l^2)$. By average $R_{0,l}$ over $\mathbb{CN}(0,\sigma_l^2)$ a closed-form derivation can be obtained as

$$R_{0,l} = 1 - \frac{1}{2\beta\sigma_l^2}E_1\left(\frac{\sigma_n^2}{2\beta\sigma_l^2}\right)\sigma_n^2 e^{\frac{\sigma_n^2}{2\beta\sigma_l^2}} \quad (20)$$

where $E_1(x) = \int_1^\infty e^{-xt}/t\,dt$. Based on (20), the rate loss due to the use of finite size codebooks increases with increased overall receiving SNR. On the other hand, the second multiplication term $R_{1,l}$ in (19) depends only on the wavenumber errors $\kappa'_{l,BS(UE)}$ and the number of antenna elements. Again, assuming $\kappa'_{l,BS(UE)}$ is small enough, $\left|\Delta_{l,BS(UE)}^2\right|$ in (17) can be approximated to be

$$\left|\Delta_{l,BS(UE)}^2\right| \approx \left|1 - (N_{BS(UE)}^2 - 1)(d)^2\kappa'^2_{l,BS(UE)}/24\right|^2 \quad (21)$$

Here $\left|\Delta_{l,BS(UE)}^2\right|$ decreases with increasing absolute value of wavenumber error $\left|\kappa'_{l,BS(UE)}\right|$. According to (10) and (13), the maximal wave number error reads $\left|\kappa'_{l,BS(UE),\max}\right| = 2\pi/C_{BS(UE)}/\lambda$. By setting $\left|\Delta_{l,BS(UE)}^2\right| \geq 0.5$ and $\left|\kappa'_{l,BS(UE)}\right| = \left|\kappa'_{l,BS(UE),\max}\right|$ in (21), the minimal codebook size for avoiding instantaneous 3 dB power loss of $\left|\Delta_{l,BS(UE)}^2\right|$ reads

$$C_{BS(UE)}^2 \geq 0.57\pi^2\left(\frac{d}{\lambda}\right)^2\left(N_{BS(UE)}^2 - 1\right) \quad (22)$$

Assuming the numbers of antennas at the BS and UE are not small and antenna distance is set to be $d = \lambda/2$, a simple design rule of thumb between the number antennas and proper codebook size can be obtained by evaluating

$$C_{BS(UE)} > 1.18N_{BS(UE)} > N_{BS(UE)} \quad (23)$$

The formula in (23) shows, in order to avoid larger than 3 dB instantaneous power loss for each path, we should use a code-book size that is at least larger than the number of antenna elements implemented in the array at the BS and UE side.

In addition, as shown in (13) and assuming $d = \lambda / 2$, we have $d\kappa'_{l,BS} \approx \pi\phi' \cos\phi_l \leq \pi\phi'$ and $d\kappa'_{l,UE} \approx \pi\theta'_l \cos\theta_l \leq \pi\theta'$. The value of $R_{1,l}$ in (19) is upper bounded by the case that $d\kappa'_{BS}$ and $d\kappa'_{UE}$ follow uniform distributions as $d\kappa'_{BS} \sim [-\pi/C_{BS}, \pi/C_{BS}]$ and $d\kappa'_{UE} \sim [-\pi/C_{UE}, \pi/C_{UE}]$ respectively. Then, after some mathematical derivations, a closed-form upper bound for $R_{1,l}$ can be obtained as

$$R_{1,l} \approx 1 - \left(1 + \frac{\gamma_{l,BS}^4 a_{BS}^2}{5} - \frac{2\gamma_{l,BS}^2 a_{BS}}{3}\right) \times \left(1 + \frac{\gamma_{l,UE}^4 a_{UE}^2}{5} - \frac{2\gamma_{l,UE}^2 a_{UE}}{3}\right) \quad (24)$$

where $a_{BS(UE)} = (\pi d / \lambda)^2 / 6$ and $\gamma_{l,BS(UE)} = N_{BS(UE)} / C_{BS(UE)}$.

Based on (19), (20) and (24), rate loss increases with increasing SNR per path and also with increasing ratio between the number of antenna elements and codebook size $\gamma_{l,BS(UE)}$.

## IV. NUMERICAL RESULTS AND ANALYSIS

In this section, we study performance of the proposed analog beamsteering approach using extensive computer simulations. Assume the system operates at central frequency of 28 GHz. There are $L = 3$ paths in the propagation environment with path-loss exponent factor of 2. The gain of each path is modeled as complex Gaussian distributed variable following $\mathbb{CN}(0, 1/L)$. The transmitted power is 27 dBm and the distance between BS and UE is set to be 50 m. We further assume 8 antennas are deployed at the UE side. For comparison purpose, all the SNRs depicted in the figures are defined as $\chi = P_s \sum_{l=1}^{L} \sigma_l^2 / (\rho \sigma_n^2)$.

First, the performance of the proposed analog beamsteering approach using a codebook with infinite precision is compared with SVD based digital beamforming. 8 or 64 antennas are deployed at the BS side. As shown in Fig. 2, the proposed beamforming approach can provide maximal achievable rate very close to SVD based beamforming in both cases. It generally shows that analog beamforming alone could provide reasonable good rate performance in the low to medium SNR region where transmit/receive power is the most influential factor. The simplified rate calculation in (9) is shown to be a good approximation in the noise dominant low SNR region. CDFs of condition numbers of the effective channel matrices using digital beamforming and analog beamforming are compared in Fig. 3 using 8 and 64 antennas at the BS respectively. It shows the proposed analog beamforming has higher probability to generate ill-conditioned effective channel matrix than its digital beamforming counterpart. However if spatial multiplexing gain is only obtained when the condition number is less 100, the analog beamforming with 64 antennas at the BS can achieve performance very close to digital beamforming.

Next, we examine the impact of codebook size on the performance of the proposed analog beamforming. Assume 64 antennas are implemented at the BS and 8 antennas are implemented at the UE. The maximal achievable rates for using codebook sizes $C_{BS} = 128, 64, 16$ and $C_{UE} = 16, 8, 4$ are evaluated using extensive computer simulations. Again, the SVD based digital beamforming is used as a reference. Based on the used setups, array gain is about 27 dB. We thus consider SNR per antenna in the range of -40 dB to 0 dB where most mmwave communications are expected to operate. As shown in Fig. 4, in the case $\gamma_{BS(UE)} = N_{BS(UE)} / C_{BS(UE)}$ equals to 0.5, the performance difference between analog beamforming using a codebook with infinite precision and with finite size is very small. With increased $\gamma_{BS(UE)}$, the rate loss increases also. In Fig. 5, with fixed parameters $N_{UE} = 8$ and $C_{UE} = 16$ at the UE and received SNR per antenna -10 dB, performance of the proposed analog beamforming is evaluated using codebook size $C_{BS} = N_{BS} / \gamma_{BS}$, where $\gamma_{BS} = 0.5, 1, 2$ and $N_{BS} = 2^q$, $q = 3,...,10$, at the BS. The maximal achievable rates are evaluated using both simulations and closed-form analysis based on (19), (20) and (24). In general, the proposed analysis can roughly predict performance loss yet doesn't work properly when $\gamma_{BS(UE)} = 2$. This is due to the fact that the derivations from (19)-(24) are obtained by assuming small wavenumber errors $\kappa'_{l,BS(UE)}$. It clearly shows that in order to achieve close to the performance that SVD based digital beamforming provides, the BS(UE) should use codebook size that is twice as large as the number of antenna elements at the BS(UE). On the other hand, reducing the codebook size to be the number of implemented antennas could provide a good performance and cost trade-off.

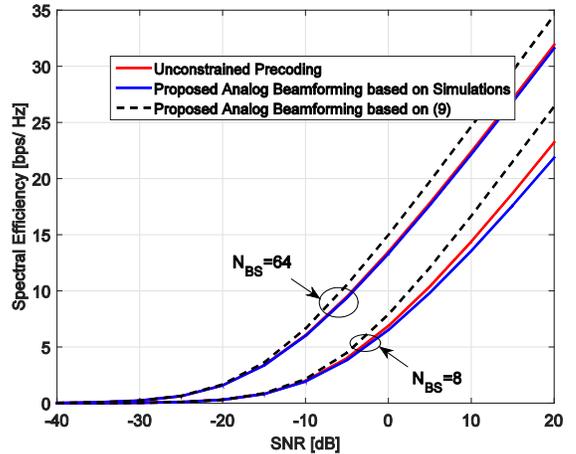

Fig. 2. Performance comparisons of proposed analog beamforming approach with SVD based digital beamforming in terms of maximal achievable rate. SNR is defined as $\chi = P_s \sum_{l=1}^{L} \sigma_l^2 / (\rho \sigma_n^2)$.

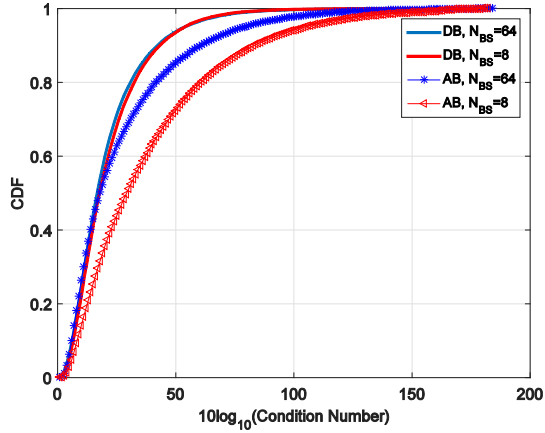

Fig. 3. Cumulative distribution probability (CDF) of $\mathrm{cond}(\mathbf{\Lambda}_L \mathbf{\Lambda}_L^H)$ and $\mathrm{cond}(\mathbf{H}\mathbf{H}^H)$, where $\mathrm{cond}(\mathbf{X})$ denotes condition number of matrix $\mathbf{X}$.

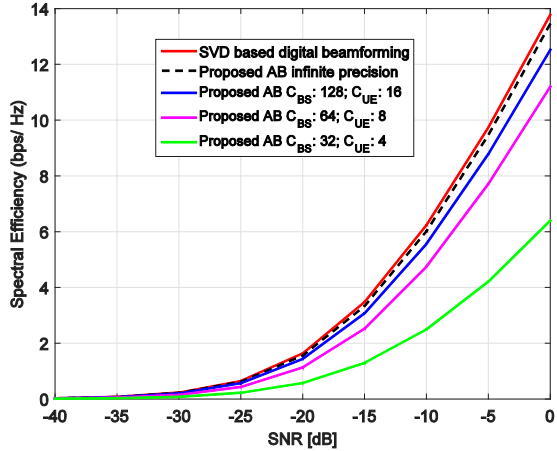

Fig. 4. Performance comparisons of SVD based digital beamforming and proposed analog beamforming (AB) using codebook with infinite precision and different codebook sizes, in terms of maximal achievable rate. SNR is defined as $\chi = P_s \sum_{l=1}^{L} \sigma_l^2 / (\rho \sigma_n^2)$.

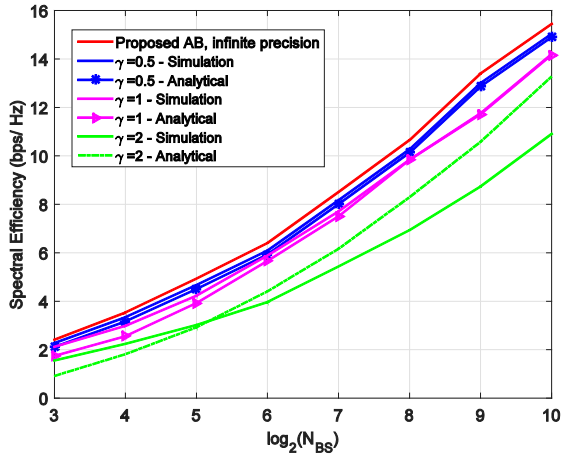

Fig. 5. Performance comparisons of proposed analog beamforming (AB) using a codebook with infinite precision and using codebook size that is $C_{BS} = N_{BS} / \gamma$ at the BS, in terms of maximal achievable rate. 8 antennas and codebook size 16 are implemented at the UE.

## V. ACKNOWLEDGEMENT

The research leading to these results received funding from the European Commission H2020 programme under grant agreement n°671650 (5G PPP mmMAGIC project).

## VI. CONCLUSIONS

In this paper, we proposed array response vector based analog beamsteering approaches that enable flexible hybrid beamforming design with performance close to SVD based digital beamforming. Assuming the use of a codebook with infinite precision at first, we construct analog beamsteering by approximating the conjugate of the array response vectors at the BS and the UE sides. The resulting effective channel matrix including both propagation channel and analog beamformers can provide maximal achievable rate very close to SVD based digital beamforming at low and medium SNR. A closed-form derivation was obtained for mapping performance degradation to the number of antennas, codebook sizes and SNR. The analysis generally showed that in order to minimize performance degradation, the codebook size should be twice as large as the number of antennas. By reducing codebook size to the number of antenna elements, a good performance and cost trade-off can be achieved. Future work will focus on devising efficient link-level and channel estimation algorithms for reaching the potential of the proposed analog beamformer in hybrid beamforming design.